\documentclass[12pt]{article}
\DeclareMathAlphabet{\scr}{U}{rsfs}{m}{n}
\usepackage{latexsym,url}
\usepackage[mathscr]{eucal}
\usepackage{amsfonts}
\usepackage{amscd}
\usepackage{cite}         
\usepackage{amssymb}
\usepackage{colordvi}
\usepackage[centertags]{amsmath}
\usepackage{enumerate}
\usepackage{graphicx}
\usepackage{booktabs}

\newcommand{\cleqn}{\setcounter{equation}{0}}
\newcommand{\newc}{\newcommand}
\newc{\be}{\begin{equation}}
\newc{\ee}{\end{equation}}
\newc{\bea}{\begin{eqnarray}}
\newc{\eea}{\end{eqnarray}}
\newc{\ol}{\overline}
\newc{\wt}{\widetilde}
\newc{\bs}{\boldsymbol}
\newc{\m}{\mathcal}

\newc{\vev}[1]{{\langle #1 \rangle}}

\newc{\mtext}{\Green}

\begin{document}

\title{\hfill ~\\[-20mm]
          \hfill\mbox{\small IPPP/13/38}\\[-3.5mm]
          \hfill\mbox{\small DCPT/13/76}\\[20mm]
       \textbf{Trimaximal TM${\bf{_1}}$ neutrino mixing in $\bs{S_4}$ \\[3mm] with spontaneous CP violation}}
\date{}
\author{\\
Christoph Luhn\footnote{E-mail: {\tt christoph.luhn@durham.ac.uk}}
\\[3mm]
  \emph{\small Institute for Particle Physics Phenomenology, University of Durham,}\\
  \emph{\small Durham, DH1 3LE, United Kingdom}}

\maketitle

\begin{abstract}
\noindent  
The measurement of the reactor angle by the Daya Bay and RENO
experiments in 2012 has ruled out the tri-bimaximal paradigm. 
Adopting an $S_4$ family symmetry, we propose direct models of the trimaximal
type TM$_1$  in which the tri-bimaximal Klein symmetry of the neutrino sector
is broken to a residual $Z_2$ symmetry. In such a scenario, the solar mixing
angle is decreased compared to its tri-bimaximal value by about $1^\circ$, thus
bringing it in excellent agreement with experimental observation. The
atmospheric mixing angle, on the other hand, depends on the CP violating Dirac phase
$\delta$. Imposing CP conservation in the family symmetry limit, we show how
to break the CP symmetry via flavon VEVs with well-defined complex phases,
so that sizable deviations of the atmospheric angle from maximal mixing,
consistent with the latest global fits, are produced. 
\end{abstract}
\thispagestyle{empty}
\vfill
\newpage
\setcounter{page}{1}

%%%%%%%%%%%%%%%%%%%%%%%%%%%%%%%%%%%%%%%%%

%%%%%%%%%%%%%%%%%%%%%%%%%%%%%%%%%%%%%%%%%

%%%%%%%%%%%%%%%%%%%%%%%%%%%%%%%%%%%%%%%%%

\section{Introduction}
\cleqn

In spring 2012, the Daya Bay~\cite{An:2012eh} and RENO~\cite{Ahn:2012nd}
experiments independently measured the reactor angle $\theta_{13}$, the
smallest mixing angle of the Pontecorvo-Maki-Nakagawa-Sakata (PMNS) matrix, to
be around $9^\circ$.  This remarkable discovery came as a surprise to many
physicists who shared a certain degree of prejudice for a smaller angle (as it
had happened previously with the solar mixing angle $\theta_{12}$). 
The paradigm of tri-bimaximal mixing~\cite{Harrison:2002er,Harrison:2002kp,Xing:2002sw,Harrison:2003aw}, which had undeniably dominated the flavour
model building landscape, was suddenly overthrown by experimental facts.
With it, the whole idea of an underlying family symmetry governing the
pattern of fermion masses and mixings was called into question. While neutrino
mixing anarchy~\cite{Hall:1999sn,Haba:2000be,deGouvea:2012ac} can
qualitatively explain large mixing angles, it fails to make quantitative and
testable predictions. On the other hand, the family symmetry
approach~\cite{Altarelli:2010gt,Ishimori:2010au,Grimus:2011fk} -- modified to
accommodate sizable $\theta_{13}$~\cite{King:2013eh} -- allows to construct
predictive models which can be falsified by future experimental data. 

In general, non-Abelian family symmetries with triplet representations allow
to unify the three chiral families. In order to realistically describe the
structure of their masses and mixings, the family symmetry needs to be
broken. This is typically achieved by means of Standard Model neutral flavon
fields which acquire  vacuum expectation values (VEVs) in appropriate
directions in flavour space. Depending on these flavon alignments, the family
symmetry approach gives rise to particular neutrino mixing patterns in either
a direct or an indirect way~\cite{King:2009ap}. In direct models, the
flavons appearing in the neutrino sector have to break the underlying family
symmetry $\m G$ down to the residual $Z_2\times Z_2$ Klein symmetry of the
(Majorana) neutrino 
mass matrix which is associated with the desired mixing pattern. In indirect
models, the flavon VEVs break $\m G$ completely, and the special structure of
the neutrino mass matrix arises in the context of the type~I seesaw~\cite{Minkowski:1977sc,ramond-seesaw,yanagida-seesaw,mohapatra-seesaw} with
sequential dominance~\cite{King:SD-a,King:SD-b,King:SD-c,King:SD-d,King:SD-e}
from the quadratic appearance of the flavon fields in the neutrino Lagrangian
or, in some rare cases, accidentally from a combination of the flavon
alignments and the group's Clebsch-Gordan
coefficients~\cite{Luhn:2012bc}. Tri-bimaximal mixing can be 
obtained in the context of constrained sequential dominance~\cite{King:2005bj}
where the required flavon alignments can be readily derived from an underlying
non-Abelian family symmetry as e.g. $\Delta(27)$~\cite{deMedeirosVarzielas:2006fc}, $Z_7\rtimes Z_3$~\cite{Luhn:2007sy} and $A_4$~\cite{deMedeirosVarzielas:2005qg,King:2006np,Antusch:2011sx}.
With tri-bimaximal mixing being ruled out by the measurement of
$\theta_{13}\approx 9^\circ$, ``non-standard'' and somewhat more complicated flavon
alignments have to be considered, leading to new and predictive versions of
constrained sequential
dominance~\cite{Antusch:2011ic,Antusch:2013wn,King:2013iva,King:2013xba}. 

In direct models, there are in principle two possible ways of generating a
mixing pattern which deviates from tri-bimaximal mixing. The first is solely
based on symmetry arguments and requires to consider larger symmetry groups
which contain a $Z_2\times Z_2$ Klein symmetry different from the
tri-bimaximal one~\cite{Toorop:2011jn,deAdelhartToorop:2011re,Ding:2012xx,Ishimori:2012gv,King:2012in,Lam:2012ga,Holthausen:2012wt,Lam:2013ng,King:2013vna}.
In this paper, however, we will pursue the second option where a tri-bimaximal setup
is augmented by an additional ingredient which breaks the tri-bimaximal structure in a
well-defined and controlled way.\footnote{A related approach adopts
non-Abelian groups which contain only half the Klein symmetry of the neutrino
sector~\cite{Ge:2011ih,Ge:2011qn,Hernandez:2012ra,Hernandez:2012sk}.}

As was shown in~\cite{Lam:2008rs,Lam:2008sh}, the natural symmetry
of tri-bimaximal mixing is the permutation group $S_4$, or any group
containing it as a subgroup (e.g. $PSL_2(7)$~\cite{Luhn:2007yr,King:2009mk,King:2009tj}). Starting with the family symmetry $S_4$, it is well known how to construct
models of tri-bimaximal neutrino mixing, see
e.g.~\cite{Bazzocchi:2008ej,Ishimori:2008fi,Bazzocchi:2009pv,Hagedorn:2010th}.  
In these models, the flavon fields appearing in the neutrino sector break
$S_4$  down to the tri-bimaximal Klein symmetry $Z_2^S \times Z_2^U$,
generated by the order-two elements $S$ and $U$. The flavons of the charged
lepton sector, on the other hand, break $S_4$ to a residual $Z_3^T$ symmetry,
corresponding to the order-three element $T$. Reversely, the three elements
$S$, $U$ and $T$ generate the group $S_4$. Their explicit matrix form depends
on the basis chosen for the five irreducible representations. It is convenient
to work in a basis with diagonal $T$ generator as this automatically yields
a diagonal charged lepton mass matrix. Our choice of basis is summarised in
Table~\ref{tab:basis}, and corresponds to the basis used
e.g. in~\cite{King:2011zj}. In this basis, the Clebsch-Gordan coefficients are
real, and we refer the reader to Appendix~A of~\cite{King:2011zj} for their
explicit values. 
\begin{table}[t]
\begin{center}
$$
\begin{array}{c|cc|c}\toprule
S_4 &  S & U & T  \\\midrule
{\bf 1,1'} &  1&\pm 1&1  \\[2mm]
{\bf 2}  & 
\begin{pmatrix} 1 & 0 \\ 0&1 \end{pmatrix} 
& \begin{pmatrix} 0& 1 \\ 1&0 \end{pmatrix}
&\begin{pmatrix} \omega & 0 \\ 0&\omega^2 \end{pmatrix}  
 \\[4mm]
{\bf 3,3'} & 
\frac{1}{3}\begin{pmatrix} -1 & 2&2 \\ 2&-1&2 \\2&2&-1 \end{pmatrix} 
& \mp \begin{pmatrix} 1&0&0\\0&0& 1 \\ 0&1&0 \end{pmatrix} 
&\begin{pmatrix} 1&0&0\\ 0&\omega^2 &0 \\ 0&0&\omega \end{pmatrix}  
\\\bottomrule
\end{array}
$$
\end{center}
\caption{\label{tab:basis}The matrix representation of the $S_4$ generators
  $S$, $U$ and $T$ for the five irreducible representations in the basis with
  diagonal $T$. Here $\omega=e^{2\pi i/3}$.}
\end{table}

A simple way of generating deviations from tri-bimaximal mixing is provided
by adding at least one extra term in the neutrino sector which does not share
the $Z_2^S\times Z_2^U$ Klein symmetry, see
also~\cite{Hall:2013yha}. However, in order to retain some 
degree of predictivity, we consider only cases with a residual $Z_2$
symmetry. Out of the three possible cases, one ($Z_2^U$) forces $\theta_{13}$
to vanish, while the other two ($Z_2^S$ and $Z_2^{SU}$) allow to dial a
sizable reactor angle. Direct models featuring a remnant $Z_2^S$ symmetry have
been studied extensively in the literature, e.g.~\cite{He:2006qd,Ishimori:2010fs,Shimizu:2011xg,Ma:2011yi,King:2011zj,Cooper:2012wf,Hagedorn:2012ut,Varzielas:2012ai}, 
and arise straightforwardly in models with an $A_4$ (obtained from $S_4$ by
dropping the $U$ generator~\cite{Luhn:2007uq,Escobar:2008vc}) family
symmetry. They are known to lead to the trimaximal
TM$_2$~\cite{Albright:2010ap,He:2011gb} neutrino mixing pattern in 
which the solar angle $\theta_{12}$ retains its tri-bimaximal value to first
approximation, while second order corrections lead to a slightly larger
angle. On the other hand, direct models with a remnant $Z_2^{SU}$ symmetry
have not received a great deal of
attention~\cite{Varzielas:2012pa,Grimus:2013rw}, despite the fact that the
predicted solar angle shows better agreement with data. This case leads to the 
trimaximal TM$_1$~\cite{Albright:2010ap,He:2011gb,Rodejohann:2012cf}  neutrino
mixing pattern which is characterised by second order corrections to
$\theta_{12}$ yielding a slightly smaller solar angle compared to the
tri-bimaximal case. In this paper, we propose direct models of trimaximal
TM$_1$ mixing which automatically predict a solar angle in excellent
agreement with the data and allow to fit the reactor angle to its measured
value of about $9^\circ$. 

Due to the breaking of the neutrino Klein symmetry to a remnant $Z_2$ in the
neutrino sector, correlations between the mixing parameters of the PMNS matrix
ensue. In the physically interesting cases of a residual $ Z_2^S$ or
$Z_2^{SU}$ symmetry, the linearised version of these correlations are known as
atmospheric mixing sum rules~\cite{King:2013eh} and involve the CP violating
Dirac phase~$\delta$. The fact that these sum rule predictions involve a CP
violating phase motivates us to construct models which are able to make
statements about the CP structure of the theory. To this end, it is natural to
impose CP conservation at energies above the family symmetry breaking
scale. The CP symmetry only gets broken spontaneously by flavon VEVs which
pick up specified complex phases. As was discussed recently, it is generally
non-trivial to define a CP transformation consistently in the presence of a
non-Abelian family symmetry~\cite{Feruglio:2012cw,Holthausen:2012dk}.
It often requires what is called a generalised (or, more appropriately, general)
CP transformation~\cite{Ecker:1981wv,Ecker:1983hz,Ecker:1987qp,Neufeld:1987wa,Grimus:1995zi,Branco:2004hu,Branco:2011zb}. 
However, the situation is rather simple in the case of an $S_4$ family
symmetry formulated in the basis of Table~\ref{tab:basis}, where the general CP
transformation maps a field $\psi$ to~\cite{Ding:2013hpa,Feruglio:2013hia}
\be
\psi (t,x) ~\stackrel{\text{CP}}{\longrightarrow} ~ \rho (g) \, \psi^\ast (t,-x)\ .
\ee
$\rho(g)$ denotes the unitary matrix representation of an $S_4$ element
$g$, and the obvious action of CP on the possible spinor indices has been
suppressed. In particular we see that the naive CP transformation with
$\rho(1)=1$ is allowed in this case. As a consequence, all coupling constants
are real in an $S_4$ model with imposed CP symmetry. Then complex phases and
with it CP violation in the Yukawa couplings can only arise from the phase structure
of the flavon VEVs. As the CP phase feeds into the correlations of the mixing
parameters caused by the residual $Z_2^{SU}$ symmetry of the neutrino sector,
the atmospheric mixing angle $\theta_{23}$ will be a function of the complex
flavon VEVs. Taking the hint for deviations from maximal $\theta_{23}$ of the
order of approximately $5^\circ$
seriously~\cite{Tortola:2012te,Fogli:2012ua,GonzalezGarcia:2012sz}, we
construct the first direct models of trimaximal TM$_1$ mixing with imposed CP
symmetry.  

The paper is organised as follows. In Section~\ref{sec:2} we revisit all three
possibilities of breaking the tri-bimaximal Klein symmetry down to a residual
$Z_2$. The available flavon vacuum configurations are collected and the
resulting predictions for $\theta_{12}$ and $\theta_{23}$ are
given. Section~\ref{sec:3}, together with Appendix~\ref{app1}, discusses the
trimaximal case TM$_1$ in detail. Requirements on the phase structure of the
input parameters (i.e. flavon VEVs) are identified and the possible neutrino
mass spectra are presented. The neutrino phenomenology of models based on the
type~II~\cite{Magg:1980ut,Schechter:1980gr,Lazarides:1980nt,Mohapatra:1980yp}
as well as the type~I
seesaw~\cite{Minkowski:1977sc,ramond-seesaw,yanagida-seesaw,mohapatra-seesaw}
is scrutinised in Section~\ref{sec:4}.  The derivation of the flavon VEV configurations,
together with their phase structure is given in Section~\ref{sec:5}. Finally,
we conclude in Section~\ref{sec:6}. 

%%%%%%%%%%%%%%%%%%%%%%%%%%%%%%%%%%%%%%%%%

%%%%%%%%%%%%%%%%%%%%%%%%%%%%%%%%%%%%%%%%%

%%%%%%%%%%%%%%%%%%%%%%%%%%%%%%%%%%%%%%%%%

\section{\label{sec:2}Residual $\bs{Z_2}$ symmetries from neutrino flavon
  VEVs}
\cleqn

In the tri-bimaximal limit, the neutrino mass matrix arises from the structure
\be
\nu\nu (
\alpha_{\bf 1} \phi_{\bf 1} 
+\alpha_{\bf 2} \phi_{\bf 2}
+ \alpha_{\bf 3'} \phi_{\bf 3'}  ) \ ,
\label{eq:tbstr}
\ee
where the neutrinos $\nu$ transform in the triplet representation  ${\bf 3}$
of $S_4$. The coupling constants $\alpha_{\bf r}$ parameterise their
interaction with the flavons $\phi_{\bf r}$ (living in the representation ${\bf
  r}$ of $S_4$), whose VEVs are aligned as
\be
\vev{\phi_{\bf 1}} = \varphi_{\bf 1} \ , \qquad
\vev{\phi_{\bf 2}} = \varphi_{\bf 2} \begin{pmatrix} 1\\1\end{pmatrix}\ , \qquad
\vev{\phi_{\bf 3'}} = \varphi_{\bf 3'} \begin{pmatrix} 1\\1\\1\end{pmatrix}\ .
\label{tbvevs}
\ee
The resulting mass matrix reads
\be
M_{\mathrm{TB}} ~=~ 
x_{\bf 1}   \begin{pmatrix} 1&0&0\\0&0&1  \\0&1&0 \end{pmatrix} 
~+~x_{\bf 2}   \begin{pmatrix} 0&1&1\\1&1&0  \\1&0&1 \end{pmatrix} 
~+~x_{\bf 3'}   \begin{pmatrix} 2&-1&-1\\-1&2&-1  \\-1&-1&2 \end{pmatrix} \ ,
\label{eq:tbmass}
\ee
where $x_{\bf r} = \alpha_{\bf r} \varphi_{\bf r}$. This is most general matrix symmetric under
the triplet matrix representation of $S$ and $U$ as given in
Table~\ref{tab:basis}, and as a consequence it is diagonalised by the
tri-bimaximal mixing matrix
\be
U_{\mathrm{TB}} ~=~ \begin{pmatrix} 
\frac{2}{\sqrt{6}} & \frac{1}{\sqrt{3}}& 0 \\
-\frac{1}{\sqrt{6}} & \frac{1}{\sqrt{3}}& \frac{1}{\sqrt{2}} \\
-\frac{1}{\sqrt{6}} & \frac{1}{\sqrt{3}}& -\frac{1}{\sqrt{2}} 
\end{pmatrix} \ .
\label{eq:tbmix}
\ee
Adding a term which breaks $Z_2^S\times Z_2^U$ to a residual $Z_2$ symmetry,
adds one new (complex) degree of freedom to the neutrino mass matrix,
\be
M ~=~ M_{\mathrm{TB}} + \Delta M \ .\label{eq:mplus}
\ee
It is straightforward to work out the form of $\Delta M$ in each of the three
cases. For residual $Z_2$ symmetries generated by $U$, $S$ and $SU$, we obtain
\be
\Delta M_U=y\begin{pmatrix}2&0& 0\\ 0&0 & -1\\ 0&-1& 0 \end{pmatrix} , ~\quad
\Delta M_S=y\begin{pmatrix}0&1& -1\\ 1&-1 & 0\\ -1&0& 1 \end{pmatrix} ,~
\quad
\Delta M_{SU}=y\begin{pmatrix}0&1& -1\\ 1&2 & 0\\ -1&0& -2 \end{pmatrix} , 
\label{eq:deltaform}
\ee
respectively. Here $y$ denotes the magnitude of the Klein symmetry breaking
contribution to the neutrino mass matrix. 
Clearly, there is some ambiguity in defining the explicit form of $\Delta M$ as
any linear combination with the $S$ {\it and} $U$ preserving tri-bimaximal
mass matrices of Eq.~\eqref{eq:tbmass} yields a structure with the same
unbroken $Z_2$ symmetry. We have chosen the form of $\Delta M$ given in
Eq.~\eqref{eq:deltaform} so that the coefficient $y$ is directly related to
the term arising from coupling $\nu\nu$ to one of the Klein symmetry breaking flavon
fields of Table~\ref{tab:patterns}, which however respect one of the three
possible $Z_2$ subgroups generated by $U$, $S$ and $SU$, respectively. Again,
the VEV alignments of Table~\ref{tab:patterns} are ambiguous in the case of the representations 
${\bf 2}$ and ${\bf 3'}$ for which there exist vacuum alignments which respect
both $S$  and $U$, so that any linear combination of such alignments
would have identical symmetry properties. For instance, in the case of the
$S$ preserving doublet alignment of Table~\ref{tab:patterns}, actually {\it any}
vacuum alignment that is different from $(1,1)^T$ breaks $Z_2^S \times Z_2^U$
down to $Z_2^S$. 

\begin{table}
\begin{center}\begin{tabular}{c|ccc}\toprule
\begin{tabular}{c}residual\\ symmetry\end{tabular} 
& $U$ &  $S$ & $SU$  \\\midrule
${\bf 1}$ & $-$ &$-$ & $-$\\[3mm]
${\bf 1'}$ & $-$ & $1$ & $-$\\[3mm]
${\bf 2}$ & $-$ & $( 1 \,,\, -1 )^T $& $-$ \\[3mm]
${\bf 3}$ & ~$( 0\,,\,1\,,\,-1 )^T $~ & ~$( 1\,,\,1\,,\,1 )^T $~ & ~$(
  2\,,\,-1\,,\,-1 )^T $~ \\[3mm]
${\bf 3'}$ & $( 1\,,\,0\,,\,0 )^T $& $-$ & $( 0\,,\,1\,,\,-1 )^T $
\\\bottomrule
\end{tabular}\end{center}
\caption{\label{tab:patterns}All available vacuum configurations which break the
 $Z_2^S\times Z_2^U$ Klein symmetry of the neutrino sector to the residual
  $Z_2$ symmetries generated by $U$, $S$ and $SU$, respectively.}
\end{table}

Neutrino mass matrices which are symmetric under a residual $Z_2$
generated by $U$, $S$ or $SU$ have the useful property that they are
diagonalised by a unitary matrix which shares one of the columns of the 
tri-bimaximal mixing matrix given in Eq.~\eqref{eq:tbmix}. For unbroken $U$,
$S$ and $SU$, these are the third, second and first columns,
respectively. This can be seen by realising that any $U$ symmetric mass matrix
has an eigenvector $(0,1,-1)^T$, while $(1,1,1)^T$ and $(2,-1,-1)^T$ are
eigenvectors of mass matrices which are symmetric under $S$ and $SU$, respectively.
Alternatively, one can apply the tri-bimaximal mixing matrix $U_{\mathrm{TB}}$
on $M$ of Eq.~\eqref{eq:mplus},
\be
M' ~=~ U_{\mathrm{TB}}^T ( M_{\mathrm{TB}} + \Delta M) U_{\mathrm{TB}}~=~
M^{\mathrm{diag}}_{\mathrm{TB}} + \Delta M' \ ,
\ee
with
\be
M^{\mathrm{diag}}_{\mathrm{TB}}~=~ 
U_{\mathrm{TB}}^T M_{\mathrm{TB}}U_{\mathrm{TB}}~=~  \begin{pmatrix} 
x_{\bf 1}- x_{\bf 2} +  3x_{\bf 3'}&0&0\\
0&x_{\bf 1}+2x_{\bf 2} &0\\
0&0 & -x_{\bf 1}+ x_{\bf 2} +  3x_{\bf  3'} \end{pmatrix} \ ,\label{eq:tbp}
\ee
and
\be
\Delta M'_U = y \begin{pmatrix} 1&\sqrt{2}&0\\ \sqrt{2}&0&0
  \\ 0&0&1  \end{pmatrix},~\quad
\Delta M'_S = \sqrt{3} \, y \begin{pmatrix} 0&0&1\\ 0&0&0
  \\ 1&0&0  \end{pmatrix},~\quad
\Delta M'_{SU} = \sqrt{6} \,y \begin{pmatrix} 0&0&0\\ 0&0&1
  \\ 0&1&0  \end{pmatrix}.\label{eq:tbpdel}
\ee
This shows that the full mixing matrix which diagonalises $M$ has the form
$U_{\mathrm{TB}}U_{ij}$, where the second factor denotes a unitary
transformation involving only generations $i$ and $j$, hence leaving one
column of the tri-bimaximal mixing matrix unchanged.

The fact that one column of the mixing matrix is exactly known, allows to
formulate two predictions in each case.
\begin{itemize}
\item \underline{$U$ symmetry.} In this case, the third column of the mixing matrix takes the form
  $(0,1,-1)/\sqrt{2}$. Adopting the PDG parameterisation, we immediately find
\be
\theta_{13} = 0^\circ \ , \qquad \theta_{23} = 45^\circ \ ,
\ee
while the solar mixing angle $\theta_{12}$ remains undetermined. 
\item \underline{$S$ symmetry.} This is the trimaximal case TM$_2$~\cite{Albright:2010ap,He:2011gb}, where the
  second column of the mixing matrix takes the form $(1,1,1)/\sqrt{3}$, and we
  obtain the relation $\sin \theta_{12} \cos\theta_{13} =
  \frac{1}{\sqrt{3}}$. To first order in~$\theta_{13}$, the resulting solar
  angle takes the tri-bimaximal value of about $35.3^\circ$. However, the
  second order correction yields a shift to slightly larger angles. Inserting the
  measured values of $\theta_{13}\approx 9^\circ$, we numerically find
\be
\theta_{12} \approx 35.8^\circ \ ,
\ee
which lies outside of the 1$\sigma$ allowed regions of all three global fits
to three neutrino
mixing~\cite{Tortola:2012te,Fogli:2012ua,GonzalezGarcia:2012sz}. In fact, this
value is only barely consistent at the 3$\sigma$ level according
to~\cite{GonzalezGarcia:2012sz}. 
Concerning the atmospheric angle, it has been shown that, to first order
in~$\theta_{13}$, the following mixing sum rule holds~\cite{King:2013eh} 
\be
\theta_{23} \approx 45^\circ -\mbox{$\frac{1}{\sqrt{2}}$} \, \theta_{13} \cos \delta \ ,
\ee
which involves the CP violating Dirac phase~$\delta$. Deviations from maximal
mixing of the order of $6^\circ$ can be obtained in the case of CP
conservation, that is if $\delta=0,\pi$.
\item \underline{$SU$ symmetry.} This is the trimaximal case TM$_1$~\cite{Albright:2010ap,He:2011gb,Rodejohann:2012cf}, where the
  first column of the mixing matrix is of the form $(2,-1,-1)/\sqrt{6}$, and 
  we obtain the relation $\cos \theta_{12} \cos\theta_{13} =
  \sqrt{\frac{2}{3}}$. To first order in~$\theta_{13}$, the resulting solar
  angle again retains its tri-bimaximal value. However, this time, the second
  order correction shifts its value to slightly smaller angles. Numerically,
  using $\theta_{13}\approx 9^\circ$, we get 
\be
\theta_{12} \approx 34.2^\circ \ ,
\ee
which falls inside the 1$\sigma$ allowed regions
of~\cite{Tortola:2012te,Fogli:2012ua,GonzalezGarcia:2012sz} with the
exception of the ``free flux'' fit in~\cite{GonzalezGarcia:2012sz}, where the
1$\sigma$ region is just narrowly missed.
Similar to the case with conserved $S$ symmetry, the atmospheric angle
satisfies a mixing sum rule which now reads~\cite{King:2013eh}
\be
\theta_{23} \approx 45^\circ + \sqrt{2} \, \theta_{13} \cos \delta \ .\label{tm1sumrule}
\ee
In order to generate deviations from maximal mixing which are of the order
of $6^\circ$ a non-trivial CP phase is required, with $|\cos \delta\,| \approx
0.5$ and therefore $\delta\approx \pm 120^\circ$ for solutions in the first
$\theta_{23}$ octant and $\delta\approx \pm 60^\circ$ for solutions in the
second octant. It is interesting to note that these phases are identical to the
phases of $\omega=e^{2\pi i/3}$ and $\omega^2$ for the first octant, while the
phases required for solutions in the second octant are identical to those of
$-\omega^2$ and $-\omega$. 
\end{itemize}

%%%%%%%%%%%%%%%%%%%%%%%%%%%%%%%%%%%%%%%%%

%%%%%%%%%%%%%%%%%%%%%%%%%%%%%%%%%%%%%%%%%

%%%%%%%%%%%%%%%%%%%%%%%%%%%%%%%%%%%%%%%%%

\section{\label{sec:3}Trimaximal TM${\bf{_1}}$ mixing and CP phases}
\cleqn

We have seen in the previous section that the most general $Z_2^{SU}$
invariant mass matrix is diagonalised by a tri-bimaximal mixing matrix
followed by an additional unitary 2-3 transformation,
\be
U_{23} ~=~ \begin{pmatrix}
1&0\\0&u_{23}
\end{pmatrix} \ ,
\ee
where we parameterise the 2-3 mixing by  $\kappa,  \vartheta\in
\mathbb R$,
\be
u_{23} ~=~ \begin{pmatrix} 
e^{-i \kappa}&0 \\
0& e^{i \kappa}\end{pmatrix}  \begin{pmatrix} 
 \cos \vartheta&\sin \vartheta \\
-\sin \vartheta&  \cos \vartheta\end{pmatrix} \ .
\ee
Then the complete matrix diagonalising $M_{SU}=M_{\mathrm{TB}} + \Delta
M_{SU}$, see Eqs.~(\ref{eq:tbmass},\ref{eq:deltaform}), takes the form 
\be
U_{\nu}~=~U_{\mathrm{TB}} U_{23}  ~=~
\begin{pmatrix}
\frac{2}{\sqrt{6}} &  \cos \vartheta\frac{e^{-i \kappa}}{\sqrt{3}}   
& \sin \vartheta \frac{e^{-i \kappa}}{\sqrt{3}}  \\
-\frac{1}{\sqrt{6}} &  \cos \vartheta\frac{e^{-i \kappa}}{\sqrt{3}}  -\sin \vartheta   \frac{e^{i\kappa}}{\sqrt{2}} 
&   \sin \vartheta\frac{e^{-i \kappa}}{\sqrt{3}}  +\cos \vartheta   \frac{e^{i \kappa}}{\sqrt{2}}\\
-\frac{1}{\sqrt{6}} &  \cos \vartheta\frac{e^{-i \kappa}}{\sqrt{3}}   +\sin \vartheta\frac{e^{i\kappa}}{\sqrt{2}}
& \sin \vartheta\frac{e^{-i \kappa}}{\sqrt{3}}   -\cos \vartheta\frac{e^{i\kappa}}{\sqrt{2}}
\end{pmatrix} .\label{eq:unu}
\ee
The reactor angle $\theta_{13}$ is obtained from the 1-3 entry. As the
parameter $\vartheta$ can be both positive and negative, we write
\be
\sin\theta_{13} ~=~ \mbox{$\frac{\mathrm{sign}\, \vartheta}{\sqrt{3}}$} \, \sin \vartheta   \ ,\label{eq:react}
\ee
where $\mathrm{sign}\,\vartheta$ denotes the sign of $\vartheta$. The
atmospheric angle is similarly given by the ratio of the 2-3 and 3-3 entries of $U_\nu$, 
\bea
\tan \theta_{23} &=& 
\left| \frac{ 
\cos \vartheta  +
 \sqrt{\frac{2}{3}}e^{-2i \kappa}  \sin \vartheta }
{\cos \vartheta
-\sqrt{\frac{2}{3}}e^{-2i \kappa}\sin \vartheta}\right| \ .
\eea
This relation can be easily solved for $\theta_{23}$, leading to the expansion 
\be
\theta_{23} ~=~ 45^\circ +\mbox{$\sqrt{\frac{2}{3}}$}\, \vartheta \:
\cos(2\kappa)  + \mathcal O(\vartheta^3)\ .\label{eq:atm2}
\ee
It is worth emphasising that this simple expression is correct up to second
order in $\vartheta$. Note, however, that $2\kappa$ is not identical to the
physical CP violating oscillation phase $\delta$. Therefore, the atmospheric
sum rule of Eq.~\eqref{tm1sumrule} only holds up to linear order in the
reactor angle. With $\theta_{13} \approx 9^\circ$, Eq.~\eqref{eq:atm2} can be
written as
\be
\theta_{23} ~\approx~ 45^\circ +12.8^\circ \cdot \mathrm{sign}\, \vartheta~
\cos(2\kappa) \ .
\ee
Deviations from maximal atmospheric mixing of the  order of about $6^\circ$
are possible if  
\be
2\kappa~\approx ~\pm 120^\circ \ , \qquad
\mathrm{or} \qquad~
2\kappa~\approx ~ \pm 60^\circ \ .
\ee
This intriguing observation motivates us to relate the phase $2\kappa$ to 
flavon fields which acquire complex VEVs with phases $\omega^0$ or $\omega^1$ or
$\omega^2$.

In a first step, we investigate how the phase $2\kappa$ arises from the input
parameters $x_{\bf 1}$, $x_{\bf 2}$, $x_{\bf 3'}$ and $y$ of the most general
$Z_2^{SU}$ invariant neutrino mass matrix $M_{SU}=M_{\mathrm{TB}} + \Delta
M_{SU}$. To this end, we apply a tri-bimaximal rotation to $M_{SU}$ yielding, see Eqs.~(\ref{eq:tbp},\ref{eq:tbpdel}) 
\be
M'_{SU} ~=~ \begin{pmatrix} 
x_{\bf 1}- x_{\bf 2} +  3x_{\bf 3'}&0&0\\
0&x_{\bf 1}+2x_{\bf 2} &\sqrt{6} y \\
0&\sqrt{6} y & -x_{\bf 1}+ x_{\bf 2} +  3x_{\bf  3'} \end{pmatrix} \ .
\label{eq:mpSU}
\ee
In order to determine $U_{23}$, we consider the complex 2-3 submatrix of $M'_{SU}$
\be
m_{SU}'~=~ \begin{pmatrix} 
x_{\bf 1}+2x_{\bf 2}&\sqrt{6} y \\
\sqrt{6} y &-x_{\bf 1}+ x_{\bf 2} +  3x_{\bf  3'} \end{pmatrix} \ , 
\ee
and diagonalise 
\be
m_{SU}'{m_{SU}'}^{\!\!\!\!\dagger}\, = 
\begin{pmatrix} 
A&B \\
B^\ast &D \end{pmatrix} , \quad \mathrm{with} ~ ~ \left\{\begin{array}{lcl}
A&\!\!=\!\!&|x_{\bf 1}+2x_{\bf 2}|^2 +6|y|^2\ ,\\[2mm]
B&\!\!=\!\!& \sqrt{6}
\Big[
(x_{\bf 1}+2x_{\bf 2}) y^\ast -y  (x_{\bf 1}- x_{\bf 2} -  3x_{\bf  3'}
  )^\ast  
\Big]
\ ,\\[2mm]
D&\!\!=\!\!&|x_{\bf 1}- x_{\bf 2} -  3x_{\bf  3'}|^2+6|y|^2  \ ,
\end{array}\right.\label{eq:Bdef} 
\ee
such that $u_{23}^T m_{SU}'{m_{SU}'}^{\!\!\!\!\dagger} \,u_{23}^\ast $ becomes real and diagonal.
This requires
\be
e^{2i\kappa} ~=~ \frac{B}{\sqrt{B B^\ast}} \ , \qquad \mathrm{and} \qquad
\tan (2\vartheta) ~=~ \frac{2\sqrt{BB^\ast}}{D-A} \ .\label{eq:Bcalc} 
\ee
A simple relation between the phase $2\kappa$ and the phases of the input
parameters can be realised in the case where one of the two terms in $B$, see
Eq.~\eqref{eq:Bdef}, dominates over the other,\footnote{We point out that
  there are other special cases where simple phase relations can be obtained. For
  instance, if $x_{\bf r},y\in \mathbb R$, then $2\kappa=0$ or $\pi$. Another such
  simple scenario would be to have real $x_{\bf r}$ and purely imaginary $y$, in
  which case one would find $2\kappa=\pm \frac{\pi}{2}$, see
  e.g.~\cite{Ding:2013hpa}.} i.e. either 
\be
(i) ~~  |x_{\bf  1}+2x_{\bf 2}| \gg |x_{\bf 1}- x_{\bf 2} -  3x_{\bf
  3'}|\ ,\quad \mathrm{or} \quad~
(ii) ~~ |x_{\bf  1}+2x_{\bf 2}| \ll |x_{\bf 1}- x_{\bf 2} -  3x_{\bf
  3'}|\ .\label{eq:scenarios}
\ee
Clearly, such a situation requires some amount of tuning, which, however, is
typically unavoidable in direct family symmetry models which accommodate
realistic neutrino masses with $\frac{\Delta m_{\mathrm{atm}}^2}{\Delta
  m_{\mathrm{sol}}^2} \approx \pm 32$~\cite{Tortola:2012te,Fogli:2012ua,GonzalezGarcia:2012sz}.
Let us dwell a little bit on the size of the singular values $M_i$ of
$M_{SU}$. Ignoring the effect of the parameter $y$ in Eq.~\eqref{eq:mpSU} and
assuming no further tuning among the input 
parameters $x_{\bf 1}$, $x_{\bf 2}$, $x_{\bf 3'}$  which would
suppress~$M_1$, we find the following approximate mass ratios
\be
M_1\,:\,M_2\,:\,M_3 ~\sim ~ \left\{ \begin{array}{ll}
1 \,:\, 1 \,:\,\epsilon   \ , & \mathrm{for~case~}(i)\ ,\\[1mm] 
1 \,:\, \epsilon \,:\,1 \ ,&   \mathrm{for~case~}(ii)\ ,
\end{array}\right.\label{eq:ratios}
\ee
with $\epsilon\ll 1$.
In the case where $M_{SU}$ corresponds directly to the  
Majorana mass matrix of the left-handed neutrinos, for instance in the context
of the type~II seesaw mechanism, the pattern of case~$(i)$ suggests an inverted
neutrino mass hierarchy, while the mass ratios of case~$(ii)$ are incompatible with
the experimental data. The situation is somewhat more model dependent if the
type~I seesaw is at work, and $M_{SU}$ corresponds to the right-handed
Majorana neutrino mass matrix. In the simplest scenario where the Dirac neutrino
Yukawa matrix $Y_\nu$ is proportional to
\be
Y_\nu ~\propto~ \begin{pmatrix} 1&0&0 \\ 0&0&1\\0&1&0 \end{pmatrix} , 
\ee
the light neutrinos will have mass ratios as in Eq.~\eqref{eq:ratios} with 
$\epsilon$ replaced by $\epsilon^{-1}$. We immediately see, that the resulting
pattern for case $(i)$ suggests a normal neutrino mass hierarchy, while the pattern
for case $(ii)$ is again not viable.

In the remainder of this paper, we therefore focus on scenarios of type $(i)$
in Eq.~\eqref{eq:scenarios}, where the phase $2\kappa$ is approximately
identical to the argument of $(x_{\bf 1}+2x_{\bf 2}) y^\ast $. The idea is
then to construct models with spontaneous CP violation in which the
parameters $x_{\bf r}$ are effectively, i.e. after absorbing an overall phase, real,
while the parameter $y$ has the phase $\omega$ or $\omega^2$. As discussed
above, such a situation will drive the atmospheric angle $\theta_{23}$ away
from its maximal value by about~$6^\circ$. The direction of this shift, into
the first or second octant,  will however depend on the signs of the involved
parameters, which remain beyond the reach of pure model building arguments.

Before discussing concrete model realisations in the following sections,
we comment on possible strategies for obtaining the alignments presented in
Table~\ref{tab:patterns}. One option would be to construct them effectively
from combining two flavons in a specific way. In the $Z_2^{SU}$ symmetric
case, one could, for instance, consider a flavon triplet $\chi_{\bf 3}$ with
alignment $(1,0,0)^T$ and multiply it with the doublet flavon $\phi_{\bf 2}$
of Eq.~\eqref{tbvevs}
to generate an effective ${\bf 3'}$ flavon with the $SU$ preserving alignment
$(0,1,-1)^T$. However, as $\vev {\chi_{\bf 3}}$ breaks $Z_2^{SU}$, the TM$_1$
scenario arises accidentally, and can be easily violated by other
contributions. Indeed, coupling $\chi_{\bf 3}$ to the flavon $\phi_{\bf 3'}$
of Eq.~\eqref{tbvevs} can generate the effective doublet alignment $(1,-1)^T$
which breaks $Z_2^{SU}$. Even though it is usually possible to construct ultraviolet
completions such that only the desirable contractions are produced, it is
generally advantageous to generate the flavon alignments in
Table~\ref{tab:patterns} directly from a suitable flavon potential. 

%%%%%%%%%%%%%%%%%%%%%%%%%%%%%%%%%%%%%%%%%

%%%%%%%%%%%%%%%%%%%%%%%%%%%%%%%%%%%%%%%%%

%%%%%%%%%%%%%%%%%%%%%%%%%%%%%%%%%%%%%%%%%

\section{\label{sec:4}Neutrino phenomenology}
\cleqn

In this section we consider supersymmetric models in which the charged
lepton mass matrix is diagonal by construction. This can be readily achieved
along the lines of already existing $S_4$ models of lepton flavour,
e.g.~\cite{King:2011zj,Ding:2013hpa}. In the neutrino sector, we employ
the tri-bimaximal flavon fields $\phi_{\bf 2}$ and $\phi_{\bf 3'}$ of
Eq.~\eqref{tbvevs} together with a $Z_2^{SU}$ preserving flavon $\wt\phi_{\bf
  3'}$ which is aligned as
\be
\vev{\wt\phi_{\bf 3'}} ~=~\wt\varphi_{\bf 3'}  \begin{pmatrix} 0\\1\\-1 \end{pmatrix} \ ,\label{eq:SU-flavon}
\ee
see Table~\ref{tab:patterns}. 
Notice that we do not include the flavon $\phi_{\bf 1}$ as this choice
proves to be a good starting point for generating an inverted (normal) neutrino mass
spectrum in the context of the type~II (type~I) seesaw.\footnote{With $x_{\bf
    1}=0$ and $y=0$ in Eq.~\eqref{eq:mpSU}, the 
requirement $(i)$ of Eq.~\eqref{eq:scenarios} yields $x_{\bf 2} \approx
-3x_{\bf 3'}$, which in turn gives $M_1 \approx M_2$.} 

Tri-bimaximal mixing is obtained from coupling the neutrinos
to $\phi_{\bf 2}$ and $\phi_{\bf 3'}$ as done in Eq.~\eqref{eq:tbstr}. Adding
in the flavon $\wt\phi_{\bf   3'}$ breaks the tri-bimaximal to the
trimaximal TM$_1$ pattern, which can either happen at the same or at higher
order. In the following we will choose the former option since numerically there is no
pronounced hierarchy between the effective parameters $x_{\bf r}$ and $y$. As will
be shown below, this is due to Eq.~\eqref{eq:react} which translates the
measured reactor angle $\theta_{13}\approx 9^\circ$ to a relatively large
value for  $|\vartheta|\approx  16^\circ$. With this assumption, the coupling of
the flavon fields to the neutrinos would take the form 
$$
\nu\nu (\alpha_{\bf 2} \phi_{\bf 2} +\alpha_{\bf 3'} \phi_{\bf 3'} 
+ \wt \alpha_{\bf 3'} \wt \phi_{\bf 3'} 
)  \ .
$$ 
As such a structure would require identical quantum numbers for
$\phi_{\bf 3'}$ and $\wt\phi_{\bf 3'}$ it is not possible to distinguish these
two flavons. As a consequence, the desired flavon alignment could not be achieved
by means of symmetries alone.\footnote{We note, however, that the flavons
could be separated in an extra dimensional setup.} In order to avoid this
conclusion, we introduce the $S_4$ singlet flavons $\xi_{\bf 1}$ and  $\wt
\xi_{\bf 1}$ (with VEVs $\xi$ and $\wt \xi$, respectively), and modify the
coupling of the flavons to the neutrinos to  
\be
\nu\nu \left[ ( \alpha_{\bf 2} \phi_{\bf 2} +\alpha_{\bf 3'} \phi_{\bf 3'} )\xi_{\bf 1}
+  \wt \alpha_{\bf 3'} \wt \phi_{\bf 3'}\wt  \xi_{\bf 1}
\right]  \ .\label{eq:neutypeII}
\ee
This structure can be readily enforced by a $Z_3 \times Z_3'$ symmetry, where the
former distinguishes $\phi_{\bf 3'}$ and $\wt\phi_{\bf 3'}$, while the latter
forbids terms with only one flavon coupling to the neutrinos. The explicit charge
assignments of the fields in the neutrino sector are listed in
Table~\ref{tab:typeII}, including two auxiliary flavons $\theta_{\bf 3'}$ and 
$\wt\theta_{\bf 3'}$ (together with an associated $Z_6^\theta$ symmetry)
which are relevant for generating the $Z^{SU}_2$ preserving alignment of
$\wt\phi_{\bf 3'}$. Furthermore, the driving fields introduced in this setup
are shown and can be identified by their $U(1)_R$ charge of~2.  Before
discussing the resulting flavon potential in Section~\ref{sec:5}, we wish to
illustrate how Eq.~\eqref{eq:neutypeII} can give rise to phenomenologically
viable neutrino masses and mixings. 

\begin{table}
\begin{center}\begin{tabular}{c|c|ccccccc|cccccccc}\toprule
& $\nu$ 
& $\phi_{\bf 2}$ & $\phi_{\bf 3'}$ &$\xi_{\bf 1}$ 
&$\wt \phi_{\bf 3'}$  & $\wt  \xi_{\bf 1}$ 
& $\theta_{\bf 3'} $& $\wt\theta_{\bf 3'} $
& $A^\theta_{\bf 2}$  
& $A^\phi_{\bf 2}$ &  $A^\phi_{\bf 3}$ &  $A^{\phi\wt\theta}_{\bf 3}$ 
& $O^{\theta\wt\phi}_{\bf 1}$  &  $O^{\wt\theta\wt\phi}_{\bf 1}$ 
& $D_{\bf 1}$ & $D'_{\bf 1}$
\\\midrule
$S_4$ & ${\bf 3}$ 
& ${\bf 2}$& ${\bf 3'}$& ${\bf 1}$
& ${\bf 3'}$& ${\bf 1}$
& ${\bf 3'}$ & ${\bf 3'}$ 
& ${\bf 2}$ & ${\bf 2}$& ${\bf 3}$& ${\bf 3}$
& ${\bf 1}$& ${\bf 1}$
& ${\bf 1}$& ${\bf 1}$
\\[3mm]
$Z_{3}$ 
& $1$ 
& $1$ & $1$ & $0$ 
& $0$ & $1$ 
& $0$ & $0$ 
& $0$ & $1$& $1$& $2$
& $0$ & $0$ 
& $0$& $0$ \\[3mm]
$Z'_{3}$ 
& $0$ 
& $1$ & $1$ & $2$ 
& $1$ & $2$ 
& $0$ & $0$ 
& $0$ & $1$& $1$& $2$
& $2$ & $2$ 
& $0$ & $1$ \\[3mm]
$Z^\theta_{6}$ 
& $0$ 
& $0$ & $0$ & $0$ 
& $0$ & $0$ 
& $2$ & $3$ 
& $2$ & $0$& $0$ & $3$
& $4$ & $3$ 
& $0$ & $0$ \\[3mm]
$\!\!U(1)_R\!$ 
& $1$ 
& $0$ & $0$ & $0$ 
& $0$ & $0$ 
& $0$ & $0$ 
& $2$& $2$& $2$
& $2$ & $2$ & $2$ 
& $2$ & $2$ 
\\\bottomrule
\end{tabular}\end{center}
\caption{\label{tab:typeII}The particle content required in the
  neutrino sector of the $S_4\times Z_{3}\times Z'_{3}\times Z_{6}^\theta$
  models.} 
\end{table}

%%%%%%%%%%%%%%%%%%%%%%%%

%%%%%%%%%%%%%%%%%%%%%%%%

\subsection{Inverted mass hierarchy from type II seesaw}

We first present the case where  the structure of Eq.~\eqref{eq:neutypeII}
arises from a type~II seesaw, with $\nu$ representing the left-handed
neutrinos $\nu_L$. The Higgs triplet $\Delta_H$ is neutral under all
symmetries of Table~\ref{tab:typeII}, and does not play any role in the
discussion of the neutrino mixing. The light neutrino mass matrix generated
from Eq.~\eqref{eq:neutypeII} is identical to $M_{SU}$ of Section~\ref{sec:3} with
\be
x_{\bf 1}=0 \ , ~~\quad
x_{\bf 2}=\alpha_{\bf 2}\, \varphi_{\bf 2}\, \xi \,
\mbox{$\frac{\vev{\Delta_H}}{\Lambda^2}$}\
 , ~~\quad
x_{\bf 3'}=\alpha_{\bf 3'}\, \varphi_{\bf 3'}\, \xi \,
\mbox{$\frac{\vev{\Delta_H}}{\Lambda^2}$}\ , ~~\quad
y= \wt\alpha_{\bf 3'}\, \wt\varphi_{\bf 3'}\, \wt\xi\,
\mbox{$\frac{\vev{\Delta_H}}{\Lambda^2}$}  \ ,\label{eq:phatypeII}
\ee
where $\Lambda$ denotes a high mass scale.
Imposing CP conservation renders all coupling constants real,
and CP is only violated spontaneously by complex flavon VEVs. The desired
phase structure depends on the flavon potential. Suitable flavon VEVs are
those leading to identical phases for $\varphi_{\bf 2}\xi$ and $\varphi_{\bf 3'}\xi$,
while the phase of $\wt\varphi_{\bf 3'}\wt \xi$ has to be shifted relative to
these by $\pm\omega$ or $\pm\omega^2$. 
Anticipating the results of Section~\ref{sec:5}, the flavons 
can develop VEVs with phases
\be
\frac{\varphi_{\bf 3'}}{|\varphi_{\bf 3'}|} = 
\pm \frac{\varphi_{\bf 2}}{|\varphi_{\bf 2}|} = 
\pm \omega^k \ ,\qquad
\frac{\xi}{|\xi|} = 
\pm \omega^l \ ,\qquad
\frac{\wt\varphi_{\bf 3'}}{|\wt\varphi_{\bf 3'}|} = 
\pm \omega^{2l} \ ,\qquad
\frac{\wt\xi}{|\wt\xi|} = 
\pm \omega^{\wt l} \ ,\label{eq:phas}
\ee
where $k,l,\wt l =0,1,2$ and the signs depend on the undetermined signs of the
real coupling constants of the flavon potential.
The common phase of $ \varphi_{\bf 2}\xi$, and $ \varphi_{\bf 3'}\xi$  can be
absorbed by a redefinition of the neutrino fields~$\nu_L$ in
Eq.~\eqref{eq:neutypeII}. This generates the following phases for the parameters
 $x_{\bf r}$ and $y$ in Eq.~\eqref{eq:phatypeII}, namely
\be
x_{\bf r} \in \mathbb R \ , \qquad  \frac{y}{|y|} = \pm \omega^{l+\wt l-k} = \pm
\omega^m  \ ,
\ee
where we have introduced $m=0,1,2$. We now assume that Nature has chosen one
of the two CP violating cases $m=1$ or $m=2$, thus entailing a relative phase
between~$x_{\bf r}$ and~$y$ of either $\pm 120^\circ$ or $\pm 60^\circ$.  
Together with the assumption 
that $x_{\bf 2} \approx - 3 x_{\bf 3'}$, such a  phase structure leads to the
result $2\kappa \approx \pm 120^\circ$ or $2\kappa \approx \pm 60^\circ$,
cf.~Eqs.~(\ref{eq:Bdef},\ref{eq:Bcalc}). Hence, in this model, the
atmospheric angle will be shifted away from its maximal value by an angle of
about $6^\circ$, provided the second relation in Eq.~\eqref{eq:Bcalc} yields a
value of $\vartheta\approx -16^\circ$ which is consistent with a reactor angle of about
$9^\circ$ and an inverted neutrino mass hierarchy. In the limit where $x_{\bf
  2} =  - 3 x_{\bf 3'}$, in which the phase factor $e^{2i \kappa}$ is given
exactly by either $\pm \omega$ or $\pm \omega^2$,  this requirement translates to 
$$
|y| ~=~ - \mbox{$\frac{1}{\sqrt{6}}$}\, \tan (2\vartheta )  \, |x_{\bf 2}|
~\approx ~ 0.25 \, |x_{\bf 2}|  \ .
$$
The approximate values of the effective parameters are then related as
$$
|x_{\bf 2}| \,:\,|x_{\bf 3'}| \,:\,|y|  ~\sim~
3 \,:\, 1\,:\, 0.75 \ ,
$$
which, as anticipated above, does not feature any clear hierarchical
structure, and so motivates the structure of Eq.~\eqref{eq:neutypeII} where all
terms enter at the same order.

In Appendix~\ref{app1} we sketch how the three effective parameters can be
directly determined by demanding that they give rise to the physically viable
values $\theta_{13}\approx 9^\circ$, 
$\frac{\Delta m_{\mathrm{atm}}^2}{\Delta  m_{\mathrm{sol}}^2} \approx -
32$, and 
$\Delta m_{\mathrm{atm}}^2 \approx -2.43 \cdot 10^{-3}\,\mathrm{(eV)}^2$.
Fixing the phase of $x_{\bf 3'}$  at zero, we obtain
\be
x_{\bf 2}\approx -0.0228\, \mathrm{eV} \ ,\qquad
x_{\bf 3'}\approx 0.0086\, \mathrm{eV} \ ,\qquad
y \approx (-1)^p \omega^m \cdot 0.0055 \, \mathrm{eV}\ ,\label{eq:numer}
\ee
where $p=0,1$ and $m=1,2$. Having fixed these input parameters, all other
physical parameters of the neutrino sector are {\it predicted}. The four choices of
the phase of $y$, see Eq.~\eqref{eq:numer}, cause a discrete ambiguity in the
obtained mixing parameters,
\bea
\theta_{12} \approx 34.2^\circ  ,~\quad
\theta_{23} \approx 45^\circ -(-1)^p \cdot 5.7^\circ  ,~\quad
\delta \approx [1+(-1)^p]\cdot90^\circ - (-1)^m \cdot 66^\circ ,\label{eq:numer1}
\eea
or more explicitly
\bea
p=0: &&\theta_{23}\approx 39.3^\circ \ , \quad \delta\approx
\left\{\begin{array}{ll} 
- 114^\circ\ , & m=1 \ ,\\ 
+114^\circ\ , & m=2 \ , \end{array}\right.
\\
p=1: &&\theta_{23}\approx  50.7^\circ \ , \quad \delta\approx 
\left\{\begin{array}{ll} 
\:\;+ 66^\circ\ , & m=1 \ ,\\ 
\:\;- 66^\circ\ , & m=2 \ . \end{array}\right.
\eea
This shows that solutions of the atmospheric mixing angle in the first octant
are predicted for $p=0$, while the choice $p=1$ gives
$\theta_{23}>45^\circ$. We emphasise that these are predictions for
$\theta_{23}$ and $\delta$ which, for a suitable choice of $p$ and $m$, happen
to be consistent with all three global fits at the 1$\sigma$ level. The
allowed regions of the fit in~\cite{GonzalezGarcia:2012sz} are met for
$(p,m)=(0,1)$ and $(p,m)=(1,2)$, while the global fit in~\cite{Fogli:2012ua}
requires $(p,m)=(0,1)$. The analysis of~\cite{Tortola:2012te} only yields a
1$\sigma$ solution in the second octant for inverted neutrino mass ordering
and does not constrain the Dirac phase at all, hence, it can be described
consistently by the choice $(p,m)=(1,1)$ and $(p,m)=(1,2)$.
The linear sum rule of Eq.~\eqref{tm1sumrule} agrees
well with the more accurate result in Eq.~\eqref{eq:numer1}: for $p=0$
($p=1$), Eq.~\eqref{eq:numer1} gives a Dirac phase $\delta\approx\mp 114$
($\delta\approx\pm 66$), which in turn yields the linear sum rule $\theta_{23}
\approx 45^\circ-5.2^\circ$ ($\theta_{23} \approx 45^\circ+5.2^\circ$). 
The neutrino masses, on the other hand, are independent of $p$ and $m$,
\be
m_{\nu_1} \approx 0.0486 \, \mathrm{eV}  \ , \qquad
m_{\nu_2} \approx 0.0494\, \mathrm{eV}  \ , \qquad
m_{\nu_3} \approx 0.0033\, \mathrm{eV}  \ . \label{eq:massIin}
\ee
Likewise, one can determine the effective mass $m_{\beta\beta}$ relevant for
neutrinoless double beta decay without ambiguity, yielding
\be
m_{\beta\beta} ~=~ \left| ({M_{SU}})_{11}\right| ~=~ 
|2 x_{\bf 3'}| ~\approx~ 0.017\, \mathrm{eV}  \ .\label{eq:massIbeta}
\ee

%%%%%%%%%%%%%%%%%%%%

%%%%%%%%%%%%%%%%%%%%

\subsection{Normal mass hierarchy from type I seesaw}

In the case of a type~I seesaw model, the structure of
Eq.~\eqref{eq:neutypeII} arises for the right-handed neutrinos $\nu=\nu_R$.
The matrix $M_{SU}$ of Section~\ref{sec:3}, which depends on the effective
parameters
\be
x_{\bf 1}=0 \ , ~~\quad
x_{\bf 2}=\alpha_{\bf 2}\, \varphi_{\bf 2}\, \xi \,
\mbox{$\frac{1}{\Lambda}$}\
 , ~~\quad
x_{\bf 3'}=\alpha_{\bf 3'}\, \varphi_{\bf 3'}\, \xi \,
\mbox{$\frac{1}{\Lambda}$}\ , ~~\quad
y= \wt\alpha_{\bf 3'}\, \wt\varphi_{\bf 3'}\, \wt\xi\,
\mbox{$\frac{1}{\Lambda}$}  \ ,\label{eq:phatypeI}
\ee
then corresponds to the right-handed neutrino mass matrix $M_R$. In order to
find the effective light neutrino mass matrix, we need to fix the structure of
the Dirac Yukawa coupling. In the simplest case, the $S_4\times Z_3\times
Z'_3\times Z^\theta_6$ charge assignments of the lepton doublet $L$ are chosen
such that they allow for the trivial coupling $y_D L\nu_R H_u$, with $H_u$
denoting the (flavour blind) up-type Higgs doublet. The Dirac neutrino mass
matrix then takes the form
\be
m_D ~=~ y_D v_u \begin{pmatrix}1&0&0\\0&0&1\\0&1&0 \end{pmatrix} ,\label{eq:trivD}
\ee
where $v_u$ is the VEV of $H_u$.
Application of the seesaw formula yields the light neutrino mass matrix
\bea
m_\nu^{\mathrm{eff}} ~=~
m_D
M_{SU}^{~-1}
m_D^T
~=~ (y_D v_u)^2 
\begin{pmatrix}1&0&0\\0&0&1\\0&1&0 \end{pmatrix}
U_\nu\, ({M^{\mathrm{diag}}_{SU}})^{-1}\,U_\nu^T
\begin{pmatrix}1&0&0\\0&0&1\\0&1&0 \end{pmatrix},
\eea
where $U_\nu$ is given in Eq.~\eqref{eq:unu} and can be obtained as discussed
in Section~\ref{sec:3}. Knowing $U_\nu$ one trivially finds the unitary
matrix which diagonalises $m_\nu^{\mathrm{eff}}$, that is the PMNS mixing
matrix
\be
U_{\mathrm{PMNS}} ~=~ \begin{pmatrix}1&0&0\\0&0&1\\0&1&0 \end{pmatrix}
U_\nu^\ast \ .\label{eq:pmnstI}
\ee 
The physical mixing parameters are therefore identical to those of $U_\nu$,
with the exception of the 2-3 mixing angle which changes the octant
$\theta_{23} \rightarrow 90^\circ -\theta_{23}$ as well as the CP
phase~$\delta$ for which $\delta\rightarrow -\delta + \pi$. The light 
neutrino masses $m_{\nu_i}$ are given by the singular values $M_i$ of $M_{SU}$, see
Eqs.~(\ref{eq:eig1}-\ref{eq:eig3}), and the product $y_D v_u$,
\be
m_{\nu_i} ~=~ \frac{(y_Dv_u)^2}{M_i} \ .\label{eq:masstypII}
\ee
Analogous to the type~II setup, the relative size of the input parameters
$x_{\bf 2}$, $x_{\bf 3'}$ and $y$ is determined by requiring
$\theta_{13}\approx 9^\circ$ and 
$\frac{\Delta m_{\mathrm{atm}}^2}{\Delta  m_{\mathrm{sol}}^2} \approx + 32$,
while their absolute magnitude cannot be fixed uniquely due to the factor
$(y_Dv_u)^2$ in Eq.~\eqref{eq:masstypII}. Demanding
$\Delta m_{\mathrm{atm}}^2 \approx 2.47 \cdot 10^{-3}\,\mathrm{(eV)}^2$, we obtain
\be
\frac{x_{\bf 2}}{(y_Dv_u)^2} \approx -32.74 \, (\mathrm{eV})^{-1} ,~\quad
\frac{x_{\bf 3'}}{(y_Dv_u)^2} \approx 18.33 \, (\mathrm{eV})^{-1} ,~\quad
\frac{y}{(y_Dv_u)^2} \approx (-1)^p \omega^m \cdot \,5.99 \,
(\mathrm{eV})^{-1}\ .
\label{eq:tyone}
\ee
The resulting predictions for the neutrino mixing parameters, $\theta_{12}
\approx  34.2^\circ$ and
\bea
p=0: &&\theta_{23}\approx 48.4^\circ \ , \quad \delta\approx
\left\{\begin{array}{ll} 
\:\:- 76^\circ\ , & m=1 \ ,\\ 
\:\:+76^\circ\ , & m=2 \ , \end{array}\right.
\\
p=1: &&\theta_{23}\approx 41.6^\circ \ , \quad \delta\approx 
\left\{\begin{array}{ll} 
+ 104^\circ\ , & m=1 \ ,\\ 
-104^\circ\ , & m=2 \ , \end{array}\right.
\eea
are again consistent with the global fits for $p=1$ and a suitable choice of~$m$
(at the 1$\sigma$ level for~\cite{GonzalezGarcia:2012sz,Tortola:2012te} and
at the 2$\sigma$ level for~\cite{Fogli:2012ua}).
The light neutrino masses take the values
\be
m_{\nu_1} \approx 0.0114 \, \mathrm{eV}  \ , \qquad
m_{\nu_2} \approx 0.0144\, \mathrm{eV}  \ , \qquad
m_{\nu_3} \approx 0.0510\, \mathrm{eV}  \ ,\label{eq:massIIin}
\ee
and the effective mass $m_{\beta\beta}$ of neutrinoless double beta decay
becomes
\be
m_{\beta\beta} \,=\, (y_Dv_u)^2\left| ({M_{SU}^{~-1}})_{11}\right| \,=\,
(y_Dv_u)^2
\left|\frac{(x_{\bf 2}+x_{\bf 3'})(x_{\bf 2}+3x_{\bf 3'}) - 4y^2}{2(x_{\bf 2}-3x_{\bf 3'})(x_{\bf 2}^2+3x_{\bf 2}x_{\bf 3'}-3y^2)}
\right| \,\approx\, 0.0023\, \mathrm{eV}  \ .\label{eq:massIIbeta}
\ee

%%%%%%%%%%%%%%%%%%%%%%%%%%%%%%%%%%%%%%%%%

%%%%%%%%%%%%%%%%%%%%%%%%%%%%%%%%%%%%%%%%%

%%%%%%%%%%%%%%%%%%%%%%%%%%%%%%%%%%%%%%%%%

\section{\label{sec:5}Flavon sector}
\cleqn

In the flavon sector, we have to generate both the alignments of the flavon
VEVs as well as their phases. To achieve this we adopt the $F$-term alignment
mechanism~\cite{Altarelli:2005yp,Altarelli:2005yx}. Introducing the driving fields $A^\theta_{\bf 2}$, $A^\phi_{\bf 2}$, 
$A^\phi_{\bf 3}$ and $A^{\phi\wt\theta}_{\bf 3}$ produces the alignments of
the flavons $\theta_{\bf 3'}$, $\phi_{\bf 2}$, $\phi_{\bf 3'}$ 
and~$\wt\theta_{\bf 3'}$, respectively. The alignment of the flavon
$\wt\phi_{\bf 3'}$ which breaks the tri-bimaximal pattern while respecting the
$Z^{SU}_2$  symmetry of $S_4$, uses orthogonality conditions obtained from the
driving fields $O^{\theta\wt\phi}_{\bf 1}$ and $O^{\wt\theta\wt\phi}_{\bf  1}$. 
Subsequently, the VEVs of the flavons $\phi_{\bf 2}$, $\phi_{\bf 3'}$, $\xi_{\bf 1}$, 
$\wt\xi_{\bf 1}$, $\theta_{\bf 3'}$ and~$\wt\theta_{\bf 3'}$ are driven to
non-zero values with fixed phases due to the presence of (five copies of) the
driving field~$D_{\bf 1}$. Finally, the driving field $D'_{\bf 1}$ is
responsible for fixing the VEV of the flavon $\wt\phi_{\bf 3'}$. 

With the charges listed in Table~\ref{tab:typeII}, the leading order terms of
the effective flavon superpotential are
\bea
W^{\mathrm{eff}}_{\mathrm{flavon}} &\!\!\sim\!\!& 
A_{\bf 2}^\theta (\theta_{\bf 3'} \theta_{\bf 3'}) 
+ A_{\bf 2}^{\phi} (\phi_{\bf 2}\phi_{\bf 2}+ \phi_{\bf 3'}\phi_{\bf 3'})
+A_{\bf 3}^{\phi} (\phi_{\bf 2}\phi_{\bf 3'})
\nonumber\\
&&
+A_{\bf 3}^{\phi\wt\theta} (\phi_{\bf 2}\wt\theta_{\bf 3'} + \phi_{\bf 3'}\wt\theta_{\bf 3'})
+ O_{\bf 1}^{\theta\wt\phi} (\theta_{\bf 3'} \wt\phi_{\bf 3'} )
+O_{\bf 1}^{\wt\theta\wt\phi} (\wt\theta_{\bf 3'} \wt\phi_{\bf 3'} )
\nonumber\\
&&+D_{\bf 1}\!
 \left[ 
\frac{(\phi_{\bf 2})^3 
+ (\phi_{\bf 3'})^3
+ \phi_{\bf 2}   (\phi_{\bf 3'})^2
+ (\xi_{\bf 1})^3
+ (\wt\phi_{\bf 3'})^3
+ (\wt\xi_{\bf 1})^3
+  (\theta_{\bf  3'})^3 }{\Lambda} 
+(\wt\theta_{\bf 3'})^2
-M^2 \right]
\nonumber\\
&&+D'_{\bf 1} \left[ (\wt\phi_{\bf 3'})^2-M \xi_{\bf 1}  \right] \, ,
\label{eq:fpot}
\eea
where we have suppressed the dimensionless coupling constants. $M$ and
$\Lambda$ denote high 
mass scales, with~$\Lambda$ being related to the mass of certain messenger fields
which generate the respective non-renormalisable operators. We emphasise that
these will generally be different for different operators. In the following we
discuss the individual terms of the flavon potential in turn.  

Starting with the driving field $A^{\theta}_{\bf 2}$, the derived $F$-term condition
reads
\be
\begin{pmatrix}
\vev{\theta_{\bf 3'}}_2^2 +2\vev{\theta_{\bf 3'}}_3\vev{\theta_{\bf 3'}}_1 \\[1mm]
\vev{\theta_{\bf 3'}}_3^2 +2\vev{\theta_{\bf 3'}}_1\vev{\theta_{\bf 3'}}_2 
\end{pmatrix} = 
\begin{pmatrix}
0\\[1mm]0
\end{pmatrix} \ .
\ee
It is straightforward to show that the most general solution to this equation
takes the form
\be
\vev{\theta_{\bf 3'}} ~=~ \theta \begin{pmatrix}1\\0\\0 \end{pmatrix} \ ,
\label{eq:align-theta}
\ee
as well as alignments obtained from this by applying any of the 24 $S_4$
transformations. With $S_4$ being a symmetry of the theory, we can choose the
alignment given in Eq.~\eqref{eq:align-theta} without loss of generality.

The $F$-term conditions obtained from the driving fields $A^\phi_{\bf 2}$ and
$A^\phi_{\bf 3}$ take the form
\bea
\begin{pmatrix} 
\vev{\phi_{\bf 2}}^2_2\\[1mm]
\vev{\phi_{\bf 2}}^2_1
\end{pmatrix} 
+ f
\begin{pmatrix} 
\vev{\phi_{\bf 3'}}^2_2+2  \vev{\phi_{\bf 3'}}_3\vev{\phi_{\bf 3'}}_1\\[1mm]
\vev{\phi_{\bf 3'}}^2_3+2  \vev{\phi_{\bf 3'}}_1\vev{\phi_{\bf 3'}}_2
\end{pmatrix} 
&=&
\begin{pmatrix}
0\\[1mm]0
\end{pmatrix} 
\label{eq:aling-phi1}
\ , \\[2mm]
\vev{\phi_{\bf 2}}_1
\begin{pmatrix}\vev{\phi_{\bf 3'}}_2\\[1mm]\vev{\phi_{\bf 3'}}_3
\\[1mm]\vev{\phi_{\bf  3'}}_1 \end{pmatrix} 
-
\vev{\phi_{\bf 2}}_2
\begin{pmatrix} \vev{\phi_{\bf 3'}}_3\\[1mm]\vev{\phi_{\bf 3'}}_1
\\[1mm]\vev{\phi_{\bf  3'}}_2 \end{pmatrix} 
&=&
\begin{pmatrix}
0\\[1mm]0\\[1mm]0
\end{pmatrix} \ , \label{eq:aling-phi2}
\eea
where a relative coupling constant $f$ has been introduced in
Eq.~\eqref{eq:aling-phi1} since this condition arises from two independent
terms in the effective flavon superpotential of Eq.~\eqref{eq:fpot}. The most
general solution to Eq.~\eqref{eq:aling-phi2} with non-zero VEVs 
$\varphi_{\bf 2}$  and $\varphi_{\bf 3'}$ is given by
$$
\vev{\phi_{\bf 2}} ~=~ \varphi_{\bf 2}
\begin{pmatrix}\omega^{2k} \\ \omega^k \end{pmatrix} \ , \qquad
\vev{\phi_{\bf 3'}} ~=~ \varphi_{\bf 3'}
\begin{pmatrix}1\\ \omega^{k} \\ \omega^{2k} \end{pmatrix} \ ,
$$
with $k=0,1,2$. Application of the $S_4$ transformation $T^k$ brings these
alignments into the standard form of Eq.~\eqref{tbvevs}. Notice that such a
$T$ transformation does not change the alignment of the flavon $\theta_{\bf 3'}$ in
Eq.~\eqref{eq:align-theta}. The VEVs $\varphi_{\bf 2}$  and $\varphi_{\bf 3'}$
are related via Eq.~\eqref{eq:aling-phi1}, yielding
\be
{\varphi_{\bf 3'}}^2 ~=~ -\,\frac{1}{3f} \, {\varphi_{\bf 2}}^2 \ .\label{eq:corre}
\ee
As we are interested in the case where the phases of $\varphi_{\bf 2}$  and
$\varphi_{\bf 3'}$ are identical (up to a possible minus sign), we impose that the
relative coupling constant $f$ be negative.

The alignment of the auxiliary flavon field $\wt\theta_{\bf 3'}$ arises due to
the $F$-term equations of $A^{\phi\wt\theta}_{\bf 3}$. Inserting the already
determined flavon alignments $\vev{\phi_{\bf 2}}$ and $\vev{\phi_{\bf 3'}}$,
see Eq.~\eqref{tbvevs}, yields
\bea
\begin{pmatrix}
\vev{\wt\theta_{\bf 3'}}_2- \vev{\wt\theta_{\bf 3'}}_3\\[1mm]
\vev{\wt\theta_{\bf 3'}}_3- \vev{\wt\theta_{\bf 3'}}_1\\[1mm]
\vev{\wt\theta_{\bf 3'}}_1- \vev{\wt\theta_{\bf 3'}}_2
\end{pmatrix} 
+ g
\begin{pmatrix}
\vev{\wt\theta_{\bf 3'}}_3- \vev{\wt\theta_{\bf 3'}}_2\\[1mm]
\vev{\wt\theta_{\bf 3'}}_2- \vev{\wt\theta_{\bf 3'}}_1\\[1mm]
\vev{\wt\theta_{\bf 3'}}_1- \vev{\wt\theta_{\bf 3'}}_3
\end{pmatrix} 
&=&
\begin{pmatrix}
0\\[1mm]0\\[1mm]0
\end{pmatrix} \ , \label{eq:aling-wttheta}
\eea
where the scales of the flavon VEVs $\varphi_{\bf 2}$ and $\varphi_{\bf 3'}$ have
been absorbed into the relative coupling constant $g$. As this constant is
generically not equal to $\pm 1$, Eq.~\eqref{eq:aling-wttheta} implies the
alignment
\be
\vev{\wt\theta_{\bf 3'}} ~=~ \wt\theta \begin{pmatrix}1\\1\\1 \end{pmatrix} \ .
\label{eq:align-ttheta}
\ee

Now that $\theta_{\bf 3'}$,  $\phi_{\bf 2}$, $\phi_{\bf 3'}$ and $\wt\theta_{\bf 3'}$
are aligned, we can derive the alignment of the flavon $\wt\phi_{\bf 3'}$
using the driving fields $O^{\theta\wt\phi}_{\bf 1}$ and 
$O^{\wt\theta\wt\phi}_{\bf 1}$. The resulting $F$-term conditions demand
orthogonality of $\vev{\wt\phi_{\bf 3'}}$ with $\vev{\theta_{\bf 3'}}$ as well as
with $\vev{\wt\theta_{\bf 3'}}$. This immediately produces the $Z_2^{SU}$
preserving alignment of Eq.~\eqref{eq:SU-flavon}. 

The third line the flavon superpotential in Eq.~\eqref{eq:fpot} generates
constraints on only the magnitude and phase of the flavon VEVs. 
Due to the imposed $Z_3$ symmetries, most flavons couple cubically to the
driving field $D_{\bf 1}$, which is neutral under all discrete
symmetries. Only the auxiliary flavon $\wt\theta_{\bf 3'}$ is allowed to
couple quadratically to $D_{\bf 1}$. Furthermore, we get exactly one mixed
term involving a product of $\phi_{\bf 2}$ and $\phi_{\bf 3'}$ flavons which
can be contracted to an $S_4$ singlet. Due to the correlation between
$\varphi_{\bf 2}$ and $\varphi_{\bf 3'}$ of Eq.~\eqref{eq:corre}, the first
three cubic terms in the third line of Eq.~\eqref{eq:fpot} can be combined
into one effective term. A further simplification is achieved by noting that
the contraction of $\vev{\wt\phi_{\bf 3'}}^3$ to an $S_4$ singlet vanishes for
the vacuum alignment of Eq.~\eqref{eq:SU-flavon}. The $F$-term condition
originating from the driving field $D_{\bf 1}$ can therefore be cast into the
simple form 
\be
\frac{h_1{ \varphi_{\bf 2}}^3  + h_2 \, \xi^3 + h_3 \, \wt\xi^3 +h_4\,\theta^3}{\Lambda} 
+ h_5 \wt\theta^2 -  M^2 ~=~ 0 \ ,\label{eq:drive}
\ee
where we have inserted all flavon VEVs and reinstated coupling constants $h_i$, which
additionally include numerical factors from the $S_4$ contractions as well as
contributions due to the combination of the first three terms in the third line
of Eq.~\eqref{eq:fpot}. Replicating this $F$-term condition five times by
introducing four more identical copies of the driving field~$D_{\bf 1}$, we
obtain a linear set of equations which allows us to decouple each term, 
\be
\frac{h'_1 \,{\varphi_{\bf 2}}^3}{\Lambda} =  M^2  \ , \qquad
\frac{h'_2\, \xi^3}{\Lambda}=  M^2  \ , \qquad
\frac{ h'_3 \, \wt\xi^3}{\Lambda}=M^2  \ , \qquad
\frac{ h'_4 \, \theta^3}{\Lambda}=M^2  \ , \qquad
h'_5 \wt\theta^2 =  M^2  \ ,\label{eq:cubes}
\ee
where $h'_i$ are new coupling constants. Thus the flavon VEVs $\varphi_{\bf2}$,
$\xi$,  $\wt \xi$, $\theta$ and $\wt \theta$ get separately driven to non-zero
values. Due to the 
imposed CP symmetry, and the resulting real coupling constants, the phases of
the first three flavon VEVs are fixed as given in Eq.~\eqref{eq:phas}. We
remark that the obtained phase predictions for the VEVs of the auxiliary
flavon fields $\theta_{\bf 3'}$ and $\wt\theta_{\bf 3'}$ do not have any
effect on the neutrino mixing parameters.  

The remaining flavon $\wt \phi_{\bf 3'}$ is driven to a non-zero VEV using 
the $F$-term equation of the driving field $D'_{\bf 1}$ in
Eq.~\eqref{eq:fpot}. Inserting the flavon VEVs 
 $\wt\varphi_{\bf 3'}$ and $\xi=\pm|\xi| \omega^l$, we find
\be
h_6 \, \wt{\varphi}_{\bf 3'}^{~2} = M |\xi| \, \omega^l \ ,\label{eq:phasetild}
\ee
where the sign ambiguity of the flavon VEV $\xi$ has been absorbed into the
coupling constant~$h_6$. In the case where the real parameters~$h_6$ and~$M$
have the same sign, the phase of~$\wt{\varphi}_{\bf 3'}$ is related to the
phase of $\xi$ as given in Eq.~\eqref{eq:phas}. With opposite signs for~$h_6$
and~$M$, an additional factor of~$i$ would arise, however, we shall not
consider this option in this paper.

%%%%%%%%%%%%%%%%%%%%%%%%%%%%%%%%%%%%%%%%%
% RENORMALIZABLE THEORY
%%%%%%%%%%%%%%%%%%%%%%%%%%%%%%%%%%%%%%%%%

Having shown how the flavon VEV configurations with the phase structure given
in Eq.~\eqref{eq:phas} can be derived from the effective flavon potential of
Eq.~\eqref{eq:fpot}, the question of higher order corrections to the flavon
alignment arises. At the purely effective level, we indeed find several
(higher) non-renormalisable terms which obey all imposed symmetries. However,
not all of these potentially dangerous terms arise in concrete ultraviolet (UV)
completions of an effective model~\cite{Varzielas:2010mp}. In such UV
completed models, the 
non-renormalisable terms of the effective theory arise by integrating out
heavy messenger fields. If no messenger field exists to mediate a
particular non-renormalisable term, this term will simply not get generated. 

In order to obtain the flavon superpotential of Eq.~\eqref{eq:fpot} it is
mandatory to introduce some messenger fields which induce the
non-renormalisable terms of the third line. All the other terms of 
$W^{\mathrm{eff}}_{\mathrm{flavon}}$ are already renormalisable, and their
existence is therefore not subject to the presence of messenger fields. 
As we have seen, there are only four non-renormalisable terms which are
relevant for driving the flavon VEVs to non-vanishing values with fixed
phases, 
\be
D_{\bf 1} \frac{1}{\Lambda} 
 \left[ (\phi_{\bf 2})^3 
+ (\xi_{\bf 1})^3 
+ (\wt\xi_{\bf 1})^3
+ (\theta_{\bf 3'})^3
  \right] \ .\label{eq:nonr}
\ee
Each of these terms requires its own pair of messenger fields denoted by $\Sigma$,
$\Sigma^c$. Their charge assignments are listed in Table~\ref{tab:mess}
\begin{table}
\begin{center}\begin{tabular}{c|cccccccc}\toprule
& $\Sigma_{\bf 2}$ & $\Sigma_{\bf 2}^c$ 
& $\Sigma_{\bf 1}$ & $\Sigma_{\bf 1}^c$ 
& $\wt\Sigma_{\bf 1}$ & $\wt\Sigma_{\bf 1}^c$ 
& $\Sigma_{\bf 3'}$ & $\Sigma_{\bf 3'}^c$ 
\\\midrule
$S_4$ & ${\bf 2}$ & ${\bf 2}$ 
& ${\bf 1}$& ${\bf 1}$
& ${\bf 1}$& ${\bf 1}$
& ${\bf 3'}$& ${\bf 3'}$
\\[3mm]
$Z_{3}$ 
& $2$ & $1$ 
& $0$ & $0$ 
& $2$ & $1$ 
& $0$ & $0$ 
\\[3mm]
$Z'_{3}$ 
& $2$ & $1$ 
& $1$ & $2$ 
& $1$ & $2$ 
& $0$ & $0$ 
 \\[3mm]
$Z^\theta_{6}$ 
& $0$ & $0$ 
& $0$ & $0$ 
& $0$ & $0$ 
& $4$ & $2$ 
 \\[3mm]
$U(1)_R$ 
& $0$ & $2$ 
& $0$ & $2$
& $0$ & $2$
 & $0$ & $2$
\\\bottomrule
\end{tabular}\end{center}
\caption{\label{tab:mess}The messenger fields required to generate (some of)
the  non-renormalisable operators of the flavon superpotential of Eq.~\eqref{eq:fpot}.} 
\end{table}
With these charges, the renormalisable superpotential involving these
messenger fields reads
\bea
W_{\Sigma} &\sim& 
D_{\bf 1} \phi_{\bf 2} \Sigma_{\bf 2}  + 
\Sigma_{\bf 2}^c  (\phi_{\bf 2} \phi_{\bf 2} +  \phi_{\bf 3'} \phi_{\bf 3'})
+\Lambda \,\Sigma_{\bf 2} \Sigma_{\bf 2}^c
\nonumber\\
&&+D_{\bf 1} \xi_{\bf 1}  \Sigma_{\bf 1}  + 
\Sigma_{\bf 1}^c \xi_{\bf 1}\xi_{\bf 1}
+\Lambda \, \Sigma_{\bf 1} \Sigma_{\bf 1}^c
\nonumber\\
&&+D_{\bf 1}\wt \xi_{\bf 1}  \wt \Sigma_{\bf 1}  + 
\wt\Sigma_{\bf 1}^c \wt\xi_{\bf 1}\wt\xi_{\bf 1}
+\Lambda \, \wt\Sigma_{\bf 1} \wt\Sigma_{\bf 1}^c \nonumber\\
&&+D_{\bf 1} \theta_{\bf 3'}  \Sigma_{\bf 3'}  + 
\Sigma_{\bf 3'}^c \theta_{\bf 3'}\theta_{\bf 3'}
+\Lambda\, \Sigma_{\bf 3'} \Sigma_{\bf 3'}^c \nonumber\\
&& + A^\phi_{\bf 2} (M\Sigma_{\bf 2})  + D'_{\bf 1} (\Sigma_{\bf 1}\Sigma_{\bf 1})
\ .\label{eq:sigma}
\eea
The first four lines of Eq.~\eqref{eq:sigma} give rise to the four effective
non-renormalisable terms of Eq.~\eqref{eq:nonr}, plus the  extra but harmless
operator $D_{\bf 1}\phi_{\bf 2}(\phi_{\bf 3'})^2 / \Lambda$,
cf. Eq.~\eqref{eq:fpot}. 
The two operators in the fifth line of Eq.~\eqref{eq:sigma} yield additional
contributions to the effective flavon superpotential,
\be
A^\phi_{\bf 2} ~ \frac{M}{\Lambda}\,(\phi_{\bf 2} \phi_{\bf 2} +  \phi_{\bf  3'} \phi_{\bf 3'}) 
~+~
D'_{\bf 1} ~\frac{{\xi_{\bf 1}}^4}{\Lambda^2} \ ,\label{eq:twoT}
\ee
which can be easily verified by systematically integrating out the $\Sigma$, $\Sigma^c$
messengers. The two terms of Eq.~\eqref{eq:twoT} arise from the
renormalisable theory involving the messengers fields of
Table~\ref{tab:mess}. We emphasise that they are the only non-renormalisable
operators which have to be added to the effective flavon superpotential
$W^{\mathrm{eff}}_{\mathrm{flavon}}$ of Eq.~\eqref{eq:fpot}. Their presence,
however,   does not change the discussion of the flavon alignment nor the
phase structure of the flavon VEVs. This can be seen by noting that the first
term of Eq.~\eqref{eq:twoT} can be absorbed into the corresponding and already
existing renormalisable term of Eq.~\eqref{eq:fpot}. The second term of
Eq.~\eqref{eq:twoT} modifies the couplings of the driving field $D'_{\bf 1}$ to
\be
D'_{\bf 1} \left[ (\wt\phi_{\bf 3'})^2-M \xi_{\bf 1} + M \xi_{\bf 1} \frac{{\xi_{\bf 1}}^3}{M\Lambda^2}  \right] .
\ee
As the cube of the VEV $\xi$ is real, see Eq.~\eqref{eq:cubes}, the resulting
$F$-term condition is of the same form as 
in Eq.~\eqref{eq:phasetild}, where the real mass parameter $M$ is slightly
corrected due to the presence of the non-renormalisable term. 
This shows that it is possible to generate the effective flavon superpotential
of Eq.~\eqref{eq:fpot} without higher order corrections other than the two
harmless operators of Eq.~\eqref{eq:twoT}. Therefore, the desired flavon
alignment, together with a particular phase structure for the flavon VEVs, can
be achieved in a UV completed model involving only a few messenger fields.  

%%%%%%%%%%%%%%%%%%%%%%%%%%%%%%%%%%%%%%%%%

%%%%%%%%%%%%%%%%%%%%%%%%%%%%%%%%%%%%%%%%%

%%%%%%%%%%%%%%%%%%%%%%%%%%%%%%%%%%%%%%%%%

\section{\label{sec:6}Conclusion}
\cleqn

The idea of an underlying family symmetry which, together with its breaking,
dictates the structure of the fermion masses and mixings has not been ruled
out by the measurement of a sizable reactor mixing angle $\theta_{13}$ of
about $9^\circ$. However, tri-bimaximal neutrino mixing (and other simple
patterns which predict vanishing $\theta_{13}$), are dead. Successful
models must necessarily involve deviations from tri-bimaximal mixing. In the
context of direct models, where the family symmetry is intimately linked to the
symmetries of the mass matrices, there exist two ways to produce new mixing
patterns. The first is based on ``large'' family symmetries which allow for
non-standard $Z_2\times Z_2$ Klein symmetries of the neutrino mass matrix. The
second approach is based on models with a tri-bimaximal $Z_2^S\times Z_2^U$
Klein symmetry which, however, gets broken to a residual $Z_2$ symmetry in
the neutrino sector. 

Assuming an underlying $S_4$ family symmetry, we have presented all available
flavon alignments which give rise to such a scenario. We focus on the
trimaximal TM$_1$ case as this is preferred over the trimaximal TM$_2$ case
due to its excellent agreement of the predicted solar mixing
angle~$\theta_{12}$ with the measured value
($\theta^{\mathrm{exp.}}_{12}\approx 34^\circ$ compared to the predictions
$\theta^{\mathrm{TM}_1}_{12}\approx 34.2^\circ$ and
$\theta^{\mathrm{TM}_2}_{12}\approx 35.8^\circ$, respectively).  Enforcing the
TM$_1$ case by means of a  remnant $Z_2^{SU}$ symmetry, we propose two
explicit supersymmetric models which respect CP symmetry in the family
symmetry limit. The CP symmetry gets broken in a controlled way when the flavon
fields acquire VEVs with well-defined complex phases. 

The first model is based on the type~II seesaw mechanism and generates an
inverted neutrino mass spectrum. Fitting the reactor angle to its measured
value, the remaining parameters of the PMNS mixing matrix are predicted within
their 1$\sigma$ experimentally allowed regions. In particular, the atmospheric
angle~$\theta_{23}$ deviates from its maximal value by about~$5.7^\circ$. The CP
violating Dirac phase~$\delta$ is predicted to be $\mp 114^\circ$ for solutions in the
first octant, while it becomes~$\pm 66^\circ$ for solutions in the second
octant. Imposing the constraints from the measured neutrino mass squared
differences, the model predicts all neutrino masses~$m_{\nu_i}$ as well as the
effective mass $m_{\beta\beta}$ of neutrinoless double beta decay, see 
Eqs.~(\ref{eq:massIin},\ref{eq:massIbeta}). 

The second model is based on the type~I seesaw and generates a normal neutrino
mass hierarchy. As in the previous model, fitting the reactor angle leads to
very good agreement of the remaining predicted mixing parameters with their
measured values. In particular, the atmospheric mixing angle~$\theta_{23}$
deviates from its maximal value by about~$3.4^\circ$. Solutions in the first octant
entail~$\delta\approx \pm 104^\circ$, while solutions in the second octant
have~$\delta\approx \mp 76^\circ$. As before, the neutrino masses
are completely fixed after matching the input parameters to 
the solar and atmospheric mass squared differences, see
Eqs.~(\ref{eq:massIIin},\ref{eq:massIIbeta}).  

To arrive at these predictions, the assumed flavon alignments have to
be justified. Here, we make use of the so-called $F$-term alignment mechanism
available in a supersymmetric context. Imposing a $U(1)_R$ symmetry as well as
two $Z_3$  and one $Z_6$ symmetry, we have studied the flavon potential in
detail. We have shown how to derive the required flavon VEV configurations as
well as the VEVs' complex phases. In order to guarantee that higher order
terms do not spoil the successful results achieved for the flavon alignments
and their phases, we have formulated a UV completion of the flavon
sector.

%%%%%%%%%%%%%%%%%%%%%%%%%%%%%%%%%%%%%%%%%

%%%%%%%%%%%%%%%%%%%%%%%%%%%%%%%%%%%%%%%%%

%%%%%%%%%%%%%%%%%%%%%%%%%%%%%%%%%%%%%%%%%

\section*{Acknowledgments}

The author thanks Gui-Jun Ding, Stephen F. King and Alexander J. Stuart for
stimulating discussions on CP symmetry in the context of family symmetry
models. He further acknowledges support from the  EU ITN grants UNILHC 237920
and INVISIBLES 289442.

%%%%%%%%%%%%%%%%%%%%%%%%%%%%%%%%%%%%%%%%%

%%%%%%%%%%%%%%%%%%%%%%%%%%%%%%%%%%%%%%%%%

%%%%%%%%%%%%%%%%%%%%%%%%%%%%%%%%%%%%%%%%%

\section*{Appendix}

\begin{appendix}

\section{\label{app1}Fixing the effective parameters $\bs{x_{\bf 2}}$, $\bs{x_{\bf
    3'}}$  and $\bs{y}$} 
\cleqn

In Section~\ref{sec:4} we have presented realisations of trimaximal TM$_1$
neutrino mixing which rely on the presence of several flavon fields. Their
VEVs give rise to three effective parameters $x_{\bf 2}$, $x_{\bf 3'}$ and
$y$. It is convenient to reparameterise these as
\be
x_{\bf 2} = a\,b \ , \qquad
x_{\bf 3'} = a \ , \qquad
y =  \omega^m \, a\,c  \ , \qquad a,b,c \in \mathbb R \ ,\label{eq:eff-P-app}
\ee
with $m=1,2$. In this notation, the entries of $m_{SU}'
{m'}_{SU}^{~~\dagger}$, see Eq.~\eqref{eq:Bdef}, can be rewritten as
\bea
A&=&a^2 \left[4b^2+6c^2  \right] \ ,\label{eq:A}\\
D&=&a^2 \left[(3+b)^2+6c^2  \right] \ ,\label{eq:D}\\
B&=& \sqrt{6} \,a^2 c \left[(3+b)\omega^m+2b\omega^{-m}  \right] \ ,\label{eq:B}
\eea
and the absolute value of $B$ simplifies to
\be
|B| ~=~ 3 \sqrt{2} \,a^2  |c| \sqrt{3+b^2}\ .\label{eq:Babs}
\ee
Notice that $A$, $D$ and $|B|$ do not depend on $m=1,2$ nor on the sign of $c$.
Plugging these expressions into the second relation of Eq.~\eqref{eq:Bcalc}
relates the real parameters $b$ and $c$ to $\tan( 2\vartheta)$,
\be
\tan( 2\vartheta) ~=~ \frac{2\sqrt{2} \sqrt{3+b^2}}{3+2b-b^2} \, |c| \ .\label{eq:elimcabs}
\ee
As the numerical value of $\vartheta$ is fixed by $\theta_{13}\approx
9^\circ$ via Eq.~\eqref{eq:react}, we obtain an expression for $|c|$ as a
function of the yet to be determined parameter $b$. In order to pin down the
physically viable value of $b$, we have to consider the neutrino masses. 
The eigenvalues of $M'_{SU} {M'}_{SU}^{~~\dagger}$ are given by
\bea
{M_1}^2 &=& a^2 (3-b)^2 \ ,\label{eq:eig1}\\
{M_2}^2 &=& A \cos^2\vartheta +D \sin^2\vartheta -|B| \sin(2\vartheta) \ ,\label{eq:eig2}\\
{M_3}^2 &=&  A \sin^2\vartheta +D \cos^2\vartheta +|B| \sin(2\vartheta)  \ .\label{eq:eig3}
\eea
In the type~II seesaw model, these masses correspond directly to the light
neutrino masses
\be
m_{\nu_i}^2 ~=~ {M_i}^2 \ .
\ee
Replacing $A$, $D$ and $|B|$ using Eqs.~(\ref{eq:A},\ref{eq:D},\ref{eq:Babs}), 
we find that the squared neutrino masses are functions of $a^2$, $b$ and
$|c|$. 
The absolute scale is fixed by the common factor $a^2$, which drops out
if we calculate the ratio of the atmospheric and the solar mass squared
differences. The parameter $|c|$ can be eliminated by means of
Eq.~\eqref{eq:elimcabs}. Then
\be
\frac{\Delta m_{\mathrm{atm}}^2}{\Delta m_{\mathrm{sol}}^2} ~=~
\frac{{M_3}^2-{M_2}^2}{{M_2}^2-{M_1}^2}\ ,
\ee
becomes a function of a single parameter, $b$, which can be calculated
numerically by setting $\frac{\Delta m_{\mathrm{atm}}^2}{\Delta
  m_{\mathrm{sol}}^2} \approx - 32$. Notice the minus sign which appears due to
the requirement of an inverse mass ordering, i.e. negative $\Delta
m_{\mathrm{atm}}^2$. Of the two possible solutions only one is consistent with
positive $\Delta m_{\mathrm{sol}}^2$. This solution requires $b\approx
-2.6515$; as a consequence of Eq.~\eqref{eq:elimcabs}, we directly get
$|c|\approx 0.6370$. Finally, the overall scale $a$ is obtained by setting $\Delta
m_{\mathrm{atm}}^2\approx -2.43\cdot 10^{-3}\,(\mathrm{eV})^2$, leading to the
value $a\approx 0.0086\,\mathrm{eV}$. The effective parameters of Eq.~\eqref{eq:eff-P-app}
are then uniquely determined up to the discrete ambiguity related to $m=1,2$
and the sign of $c$, which is parameterised by the factor $(-1)^p$ in
Eq.~\eqref{eq:numer}. 

In the type~I seesaw model, the mass matrix $M_{SU}$ of Section~\ref{sec:3}
corresponds to $M_R$, the mass matrix of the right-handed neutrinos
$\nu_R$. With a trivial Dirac mass matrix of the form as given in
Eq.~\eqref{eq:trivD}, the PMNS mixing matrix is almost identical to the
unitary matrix which diagonalises $M_{SU}$, see Eq.~\eqref{eq:pmnstI}. The
mixing angle $\vartheta$ of Eq.~\eqref{eq:elimcabs} is still related to the
reactor angle $\theta_{13}\approx 9^\circ$ via
Eq.~\eqref{eq:react}. Therefore, $|c|$ can be expressed as a function of the
parameter $b$. The latter will be fixed by considering the light neutrino
masses, which are obtained from Eqs.~(\ref{eq:eig1}-\ref{eq:eig3}) by
\be
m_{\nu_i}^2 ~=~ \frac{(y_Dv_u)^4}{{M_i}^2} \ .
\ee
As a result, the ratio of the mass squared splittings  is given by
\be
\frac{\Delta m_{\mathrm{atm}}^2}{\Delta m_{\mathrm{sol}}^2} ~=~
\frac{{M_1}^2-{M_3}^2}{{M_1}^2-{M_2}^2}  \cdot \frac{{M_2}^2}{{M_3}^2} \ .
\ee
Equating this to $+32$ for a normal neutrino mass hierarchy determines $b\approx
-1.7857$, which in turn results in $|c|\approx 0.3267$. The absolute mass
scale $a$ can finally be derived from $\Delta m_{\mathrm{atm}}^2 \approx 2.47
\cdot 10^{-3}\,(\mathrm{eV})^2$, leading to a value of $a\approx 18.33
\,\frac{(y_Dv_u)^2}{\mathrm{eV}}$.  The parameters of Eq.~\eqref{eq:eff-P-app}
are then uniquely determined for the type~I seesaw setup, up to a discrete
ambiguity which is expressed in terms of $m=1,2$ and the sign of $c$; the
latter corresponding to the factor $(-1)^p$ in Eq.~\eqref{eq:tyone}.

\end{appendix}

%%%%%%%%%%%%%%%%%%%%%%%%%%%%%%%%%%%%%%%%%

%%%%%%%%%%%%%%%%%%%%%%%%%%%%%%%%%%%%%%%%%

%%%%%%%%%%%%%%%%%%%%%%%%%%%%%%%%%%%%%%%%%

\end{document}